\documentclass[preprint,12pt]{elsarticle}

\usepackage{epstopdf}
\usepackage{amssymb}
\usepackage{amsmath}
\usepackage{graphicx}
\usepackage{natbib}
\usepackage[version=3]{mhchem}
\usepackage[colorinlistoftodos]{todonotes}
\usepackage{lineno}
\usepackage{chemformula}
\biboptions{sort&compress}
\usepackage{dcolumn}
\newcommand{\wn}{cm$^{-1}$}

\begin{document}


\title{Experimental and quantum-chemical characterization of heavy carbon subchalcogenides:\\ Infrared detection of \ch{SeC3Se}\tnoteref{t1}}
\tnotetext[t1]{This paper is dedicated to Prof. Dr. Stephan Schlemmer, on the occasion of his 60th birthday.}
%

\author[1]{Thomas Salomon}
\author[2]{John B. Dudek}
\author[2]{Yury Chernyak}
\author[3]{Jürgen Gauss}
\author[1]{Sven Thorwirth\corref{cor1}}
\ead{sthorwirth@ph1.uni-koeln.de}
\cortext[cor1]{Corresponding author}
\address[1]{I. Physikalisches Institut, Universit\"at zu K\"oln, Z\"ulpicher Str.~77, 50937 K{\"o}ln, Germany}
\address[2]{Department of Chemistry, Hartwick College, Oneonta, NY, USA}
\address[3]{Department Chemie, Johannes Gutenberg-Universit\"at Mainz, Duesbergweg 10-14, 55128 Mainz, Germany}

\begin{abstract}
High-resolution 
infrared studies of laser ablation products from carbon-se{\-}lenium targets
have revealed a new vibrational band at 2057\,\wn\ that is identified as the
$\nu_3$ vibrational fundamental of the \ch{SeC3Se} cluster.
Because of the rich isotopic composition of selenium and the heavy nuclear masses involved, the vibrational band
shows a relatively compact and complex structure despite the simple linear geometric arrangement. Overall,
rotational-vibrational lines of six isotopologues could be assigned and fitted permitting the derivation
of an accurate selenium-carbon bond length.

Spectroscopic analysis has been greatly supported by high-level quantum-chemical calculations of the molecular structure and the harmonic and anharmonic force fields performed at the CCSD(T) level of theory. Scalar-relativistic effects on the molecular structure  were also considered but found of little importance.
\end{abstract}

\begin{keyword}
Carbon subchalcogenide \sep Infrared spectroscopy \sep Laser ablation \sep CCSD(T) calculations
\end{keyword}

\maketitle

\section{Introduction}
\label{sec:intro}

Over the years, binary carbon-rich clusters C$_n$\rm X$_m$ ($n>1$, $m=1,2$) have been studied rather extensively both experimentally and theoretically
due to their relevance for molecular structure and astrochemistry \cite[see, e.g., Refs.][and references therein]{schlemmer_lab_astro,thorwirth_JPCA_120_254_2016}.
Particularly because of the latter, those clusters comprising group 14 and 16 heteroatoms, i.e., oxygen, sulfur, and silicon, have received special attention from some high-resolution spectroscopists \cite{holland_JMS_130_470_1988,mcnaughton_JMS_149_458_1991,ogata_JACS_117_3593_1995,ohshima_JCP_102_1493_1995,petry_JMSt_612_369_2002,bizzocchi_AaA_492_875_2008,tang_JMS_169_92_1995,gordon_ApJS_134_311_2001,dudek_IJQC_117_e25414_2017,thorwirth_CPL_684_262_2017,mcguire_PCCP_20_13870_2018,salomon_JMS_356_21_2019,mccarthy_JPCL_6_2107_2015,thorwirth_si2c3_2011,gordon_JCP_113_5311_2000,mccarthy_ApJ_538_766_2000,witsch_JPCA_123_4168_2019} and chains as complex as \ce{C5S} have been detected in space \cite{agundez_AAA_570_A45_2014}.
Considerably less is known for clusters containing heavier elements $\rm X$. 
Selected systems ($\rm X$=\,Ge, Ag, Cu, Ni,...) have been studied using matrix-isolation
spectroscopic techniques \cite[Refs.][and references therein]{gonzalez_JCP_130_194511_2009,szczepanski_JPCA_4778_112_2008}, however, very little data have been collected in the gas phase. 

As far as binary carbon-rich clusters with heavy group 14 elements are concerned, only very recently high-resolution studies of several germanium-bearing 
clusters have been reported, one infrared observation of nonpolar \ch{GeC3Ge}  \cite{thorwirth_JPCA_120_254_2016} as well 
as two Fourier-transform microwave investigations of polar GeC$_n$ ($n=2,4,5,6$) species 
\cite{zingsheims_JPCL_8_3776_2017,lee_PCCP_21_18911_2019}. 
The situation is even less favorable with respect to  clusters comprising carbon and
heavy group 16 elements. For selenium, only an observation of the fundamental $J=1-0$ rotational 
transition of diatomic CSe is found in the literature \cite{mcgurk_JCP_58_1420_1973} and no heavier subchalcogenides 
C$_n$Se$_m$ have been studied spectroscopically to date. So far, selected polyatomic 
carbon-selenium species
were observed using mass spectrometry, targeted at C$_n$Se$^-$ anions
in which clusters with an even number of carbon atoms as heavy as \ch{C10Se-} were detected \cite{wang_JPCA_105_4653_2001}. In organometallics, the \ce{C3Se} species has been used as a bridging ligand \cite{hill_organomet_34_361_2014}.
Some previous information on carbon-selenium species has been obtained from quantum-chemical calculations, albeit at rather modest levels of theory \cite{wang_JPCA_105_4653_2001,wu_JMStTheo_765_137_2006,bundhun_EPJD_57_355_2010,villanueva_PCCP_14_14923_2012,zhang_NJC_40_9486_2016,pu_IC_56_5567_2017}.

The present paper reports the first high-resolution spectroscopic characterization of a polyatomic carbon-selenium cluster, 
linear \ch{SeC3Se}, accomplished by observation of its antisymmetric C-C-stretching mode $\nu_3$ in the 5\,$\mu$m range. 
The experimental work was complemented by high-level quantum chemical calculations performed
at the coupled-cluster (CC) level of theory in combination with large basis sets to support the spectroscopic assignment. A possible influence of scalar-relativistic effects due to the presence of selenium 
has been evaluated. Additionally, the combination of experimental rotational constants of different isotopic species and calculated zero-point vibrational corrections permitted the determination of an accurate carbon-selenium bond length in \ce{SeC3Se}.















\section{Experimental setup}

Carbon-selenium clusters 
were observed with the same experimental setup used in recent investigations of carbon-sulfur clusters 
\cite{dudek_IJQC_117_e25414_2017,thorwirth_CPL_684_262_2017,salomon_JMS_356_21_2019}. Briefly, the spectrometer comprises a 
laser-ablation source (Nd:YAG laser frequency-tripled to operate at a wavelength of 355 nm, a repetition rate of 20 Hz, and a pulse energy of about 20 mJ) for cluster production, a widely tunable quantum cascade laser (QCL, Daylight Solutions) 
as a monochromatic radiation source and a Herriott-type multireflection cell aligned such to allow for 48 passes of the QCL beam. Sample rods were made by compressing 3:1 stoichiometric mixtures of graphite and selenium powder (Sigma-Aldrich) and a tiny amount of epoxy glue was added as binder to ensure mechanical stability.
In a running experiment, following the laser pulse, the ablation products are guided through a reaction channel (10\,mm, mounted on a Series 9 pulse valve) towards a slit exit (cross section 1\,mm $\times$ 15\,mm) using He buffer gas from a high-pressure reservoir (15 bar). At the nozzle exit, the cluster-seeded He-pulse expands adiabatically into a vacuum chamber kept at a background pressure of a few times 10$^{-2}$\,mbar resulting in typical cluster rotational temperatures of 20 to 40\,K. A few mm downstream from the nozzle exit, the QCL beam intersects the cluster pulse perpendicularly to the direction of the traveling free jet, and the transmitted laser intensity is recorded as a function of wavenumber using liquid-N$_2$-cooled InSb detectors. Frequency calibration is performed
using a wavemeter (Bristol Instruments), a Fabry-Perot \'etalon and standard calibration gases (OCS) resulting in a typical wavenumber accuracy of $\le 10^{-3}$\,\wn .  Initially, rotational-vibrational transitions of the \ce{C3} cluster \cite{matsamura_JCP_89_3491_1988} were used to optimize the experimental conditions. Later on, optimization was performed on selected transitions of \ce{SeC3Se} itself.

   
\section{Quantum-chemical calculations}





\begin{figure*}[ht!]
\centering
	\includegraphics[width=0.8\textwidth]{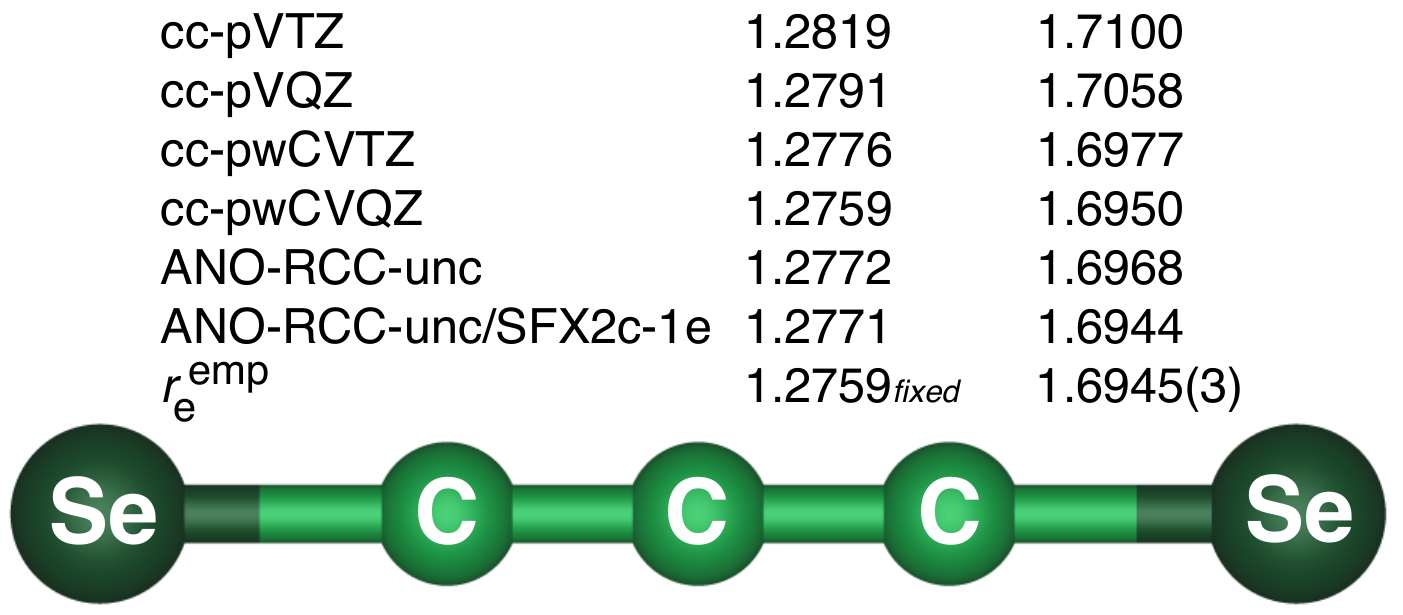}
	\caption{Structural parameters of linear \ce{SeC3Se} calculated at the CCSD(T) level of 
	theory using different basis sets as well as with and without consideration of scalar-relativistic effects (in \AA ). The semi-experimental ($r_{\rm e}^{\rm emp}$) Se$-$C bond length was determined while keeping the C$-$C distance fixed at the corresponding CCSD(T)/cc-pwCVQZ value. For further details, see text.}
	\label{se2c3struct}
\end{figure*}

The high-resolution spectroscopic study of carbon-selenium clusters reported here
was complemented with high-level quantum-chemical calculations. All calculations were performed at the
CC singles and doubles (CCSD) level augmented by a perturbative 
treatment of triple excitations, CCSD(T), \cite{raghavachari_chemphyslett_157_479_1989,shavitt_and_bartlett}
in combination with 
Dunning's 
correlation consistent polarized valence and polarized core-valence basis sets (frozen-core (fc) approximation: cc-pV$\rm X$Z; 
all-electron (ae) computations: cc-pwCV$\rm X$Z, with $\rm X$=T, Q) \cite{dunning_JCP_90_1007_1989,wilson_JCP_110_7667_1999,peterson_JCP_117_10548_2002,balabanov_JCP_123_064107_2005}) 
as well as analytic gradient techniques \cite{watts_chemphyslett_200_1-2_1_1992}.
The theoretical best-estimate structure was calculated at the ae-CCSD(T)/cc-pwCVQZ level 
of theory, an approach shown previously to provide very accurate equilibrium structural parameters for molecules
containing second-row,
\cite[e.g., Refs.][]{coriani_JCP_123_184107_2005,thorwirth_JCP_2009,mccarthy_JCP_134_034306_2011,mueck_AlCCH_2015} but also third-row main group elements such as germanium \cite{thorwirth_JPCA_120_254_2016,zingsheims_JPCL_8_3776_2017,lee_PCCP_21_18911_2019}. The structural parameters of \ce{SeC3Se} calculated at various levels of theory are summarized
in Figure \ref{se2c3struct}.

Vibrational effects were treated using second-order vibrational perturbation theory (VPT2) based on the formulas given in Ref. \cite{mills_alphas}. 
Harmonic and anharmonic force fields were calculated in the fc approximation at the CCSD(T) level of theory using basis sets as large as $\rm X=Q$
and analytic second-derivative techniques \cite{gauss_chemphyslett_276_70_1997,stanton_IntRevPhysChem_19_61_2000}
followed by numerical differentiation of the analytically computed Hessian with respect to the normal coordinates \cite{schneider_CPL_157_367_1989,stanton_JCP_108_7190_1998} to obtain the required cubic and semidiagonal quartic 
anharmonic force fields \cite{stanton_IntRevPhysChem_19_61_2000,stanton_JCP_108_7190_1998}.
These calculations provide harmonic vibrational frequencies, centrifugal distortion and vibration-rotation interaction constants, zero-point vibrational corrections to the
rotational constants $\Delta B_0$, anharmonicity constants $x_{ij}$ as well as fundamental vibrational frequencies (Table \ref{qcc_constants}) and proved instrumental in the spectroscopic assignment and analysis. 

\begin{table}[h!]
\scriptsize
\caption{Vibrational wavenumbers and rotation-vibration interaction constants of $^{80}$Se$_2^{12}$C$_3$ (vibrational modes in \wn ,
$\alpha_i$ and $q_i$ in MHz) as well as anharmonicity constants $x_{ij}$ 
relative to the $\nu_3$ mode ($x_{3j}$, in \wn ). \label{qcc_constants}}
\begin{center}
\begin{tabular}{lrrrrr|lrr|r}
\hline \hline
Vib. &
 \multicolumn{4}{c}{Calc.}& Exp. & Para- & Calc.$^a$ & Exp. & $x_{3j}$$^a$ \\ \cline{2-5}
 mode & Harm.$^a$ & Anharm.$^a$ & Harm.$^b$ & Anharm.$^c$ &  & meter \\ \hline
$\nu_1(\sigma_g)$ & 1603 &  1582 & 1606 & 1585 & $\cdots$         & $\alpha_1$       &    1.092       & $\cdots$ & $-$13.67\\
$\nu_2(\sigma_g)$ &  295 &   296 &  297 &  298 & $\cdots$         & $\alpha_2$       &    0.157       & $\cdots$ & $-$1.39 \\
$\nu_3(\sigma_u)$ & 2110 &  2070 & 2106 & 2066$^d$ & 2057.2110(1)$^e$ & $\alpha_3$   &    1.383       & 1.3656(8)$^e$ & $-$7.67 \\
$\nu_4(\sigma_u)$ &  816 &   792 &  819 &  796 & $\cdots$         & $\alpha_4$       &    0.729       & $\cdots$ & $-$3.90 \\
$\nu_5(\pi_g)$    &  362 &   355 &  359 &  352 & $\cdots$         & $\alpha_5$/$q_5$ & $-$0.417/0.026 & $\cdots$ & $-$2.77  \\
$\nu_6(\pi_u)$    &  481 &   408 &  460 &  387 & $\cdots$         & $\alpha_6$/$q_6$ & $-$0.467/0.021 & $\cdots$ & $-$11.41\\
$\nu_7(\pi_u)$    &   88 &    77 &   83 &   72 & $\cdots$         & $\alpha_7$/$q_7$ & $-$0.709/0.094 & $\cdots$ & $-$1.43  \\ \hline \hline
\end{tabular}
\end{center}
$^a$ fc-CCSD(T)/cc-pVTZ calculations.\\
$^b$ fc-CCSD(T)/cc-pVQZ calculations.\\
$^c$ Calculated from the fc-CCSD(T)/cc-pVQZ harmonic force field and anharmonic corrections calculated using VPT2 at the fc-CCSD(T)/cc-pVTZ level.\\
$^d$ Using a scaling factor derived from the $\nu_3$ mode of isovalent \ce{SC3S},
2066\,\wn\ translates into a best-estimate value of 2056\,\wn , see text.\\ 
$^e$ Gas-phase value (this work).\\
\end{table}

As in the study of the \ce{GeC3Ge} cluster \cite{thorwirth_JPCA_120_254_2016}, the possible influence of scalar-relativistic effects on the molecular structure of \ce{SeC3Se} has been explored using the spin-free exact two-component
scheme in its one-electron variant (SFX2c-1e) \cite{dyall_JCP_115_9136_2001,kutzelnigg_JCP_123_241102_2005,liu_JCP_131_031104_2009,ilias_JCP_126_064102_2007,cheng_JCP_135_084114_2011}. These calculations were performed with uncontracted versions of the ANO-RCC basis sets taken from Ref. \cite{roos_TCA_111_345_2004}.

All calculations were performed using the \textsc{Cfour} program suite \cite{cfour_JCP_2020,harding_JChemTheoryComput_4_64_2008}; a detailed review of the methods and strategies employed here can be found elsewhere \cite{puzzarini_IntRevPhysChem_29_273_2010}.

\section{Results and discussion}

\subsection{The $\nu_3$ vibrational fundamental}

Qualitatively, the appearance of the $\nu_3$ band of \ce{SeC3Se} was expected to be rather similar to the corresponding band of \ce{GeC3Ge} \cite[see Ref.][]{thorwirth_JPCA_120_254_2016}. Both clusters share similar structures and moments of inertia translating into small rotational constants
of about 350\,MHz and hence show rather compact rotation-vibration pattern. Also, both germanium and selenium have more than just one abundant stable isotope, germanium having three with natural abundances in excess of 20\% and selenium
having two, \ce{^80Se} (49.6\%) and \ce{^78Se} (23.8\%). Further selenium isotopes are found
at abundances of 9.4\% (\ce{^76Se}), 8.7\% (\ce{^82Se}),  and 7.6\% (\ce{^77Se}).
Statistical distribution of these 
isotopes over the two terminal positions in \ce{SeC3Se} gives rise to several abundant isotopic species and consequently was expected to
result in quite some spectroscopic richness in the rotational structure of the vibrational band.
As all abundant selenium isotopes as well as \ce{^12C} have no nuclear spin (i.e., $I=0$; except for \ce{^77Se} which has $I=1/2$), in addition, Bose-Einstein statistics will be at work for the symmetric species of $D_{\infty h}$ symmetry (e.g., $^{80}$SeC$_3^{80}$Se and $^{78}$SeC$_3^{78}$Se)\footnote{For the sake of simplicity, in the text isotopic species of \ce{SeC3Se} will be specified
by the corresponding selenium mass numbers only, e.g., \ce{^80SeC3^80Se} will be denoted ``80-80'' and so forth.}. 
As a consequence, line spacing in these species will be approximately $4B$ whereas the presence of two
different selenium isotopes in the same molecule will result in $C_{\infty v}$ symmetry and a regular $2B$ line spacing.
A more detailed discussion about the effect and possible consequences of spin statistics in compact and partially overlapping spectra has been given in Ref. \cite{thorwirth_JPCA_120_254_2016}.

\begin{figure*}[ht!]
\centering
    \includegraphics[width=0.8\textwidth]{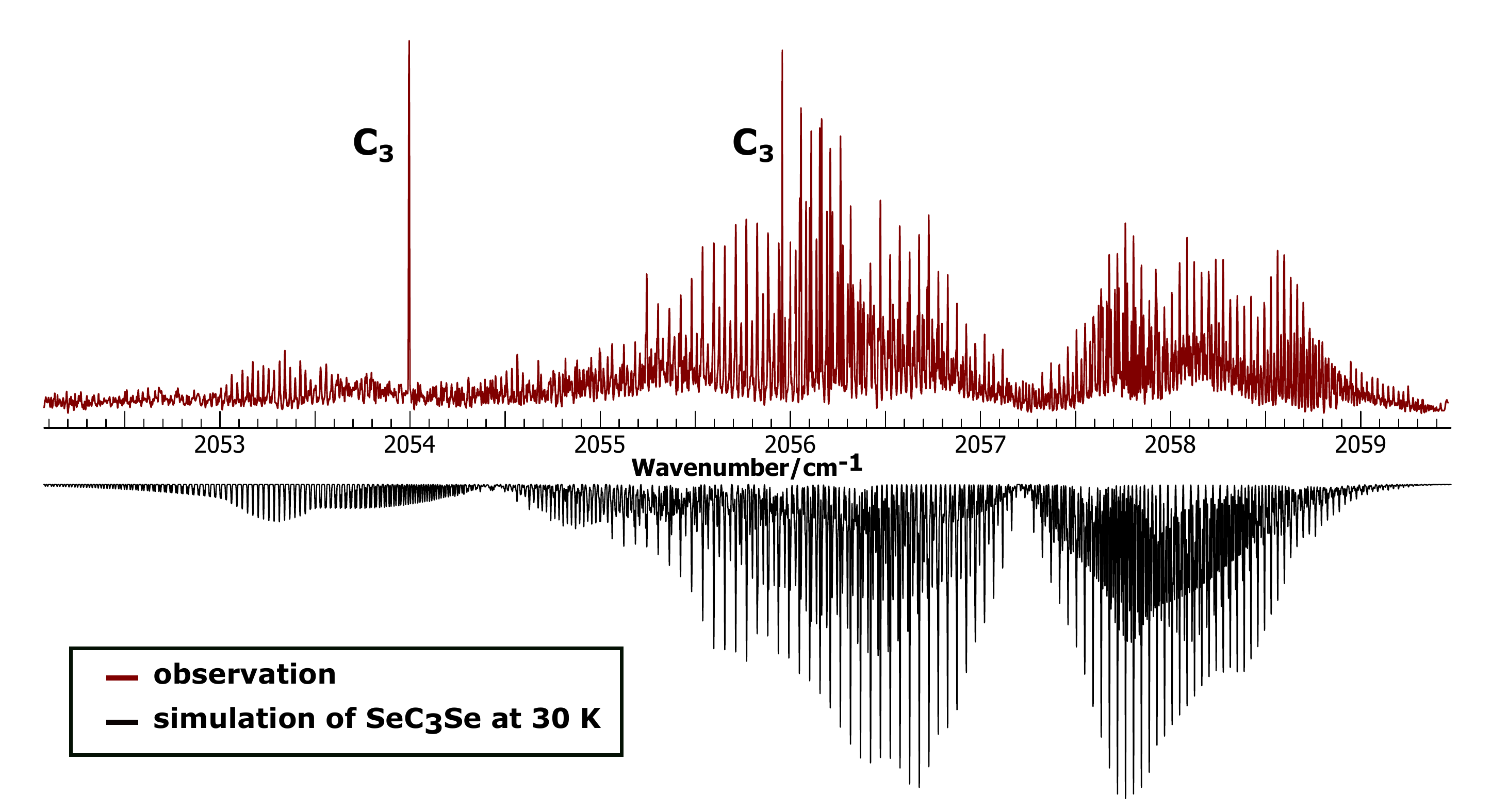}
	\caption[Experimental spectrum of the $\nu_1$ vibrational band \ce{SeC3Se}]{The $\nu_3$ vibrational band of \ce{SeC3Se} as observed in the gas phase {\it vs.} a simulation based on the best-fit parameters and a rotational temperature of 30K. In addition, two strong lines, $R$(14) at 2054\,\wn\ and $R$(16) at 2056\,\wn , of the $\nu_3$ mode of C$_3$ are prominently detected.}
	\label{overview_se2c3_nu3}
\end{figure*}

From the CCSD(T) calculations summarized in Table \ref{qcc_constants}, the location of the $\nu_3$ vibrational band of the 80--80 species 
was predicted at around 2066\,\wn . However, by comparison with the calculations and experimental spectroscopic study of structurally closely related cumulenic chains such as \ce{SC3S}, \ce{SiC3Si}, and \ce{GeC3Ge} \cite{salomon_JMS_356_21_2019,thorwirth_si2c3_2011,thorwirth_JPCA_120_254_2016} the 
location of this 
band was expected to be shifted further to the red by several \wn. More quantitatively,
using the calculated and experimental values of the $\nu_3$ band of isovalent \ce{SC3S} for calibration purposes \cite{salomon_JMS_356_21_2019}, a scaled (best estimate) value of 2056\,\wn\ is obtained for the \mbox{80--80} species.
Interestingly, by tuning the QCL to this wavenumber while running laser ablation of
a carbon-selenium rod in the very first experiment, an infrared spectroscopic signal was detected right away. Coarse tuning assays in this wavenumber range then revealed a spectroscopic pattern
expected from the calculated \ce{SeC3Se} molecular parameters
(Table \ref{v3_isotopologs})
with the spin statistical effects. Finally, repeated fine tuning of the QCL over the frequency range from 2052 to 2060\wn\ and spectral
averaging yielded the spectrum shown in Figure \ref{overview_se2c3_nu3}. The new band was only observed when target rods
comprising both carbon and selenium were used in the experiment but not through ablation from a pure carbon target alone.

\begin{table*}[ht!]
    \scriptsize
	\caption{Molecular parameters for the $\nu_3$ vibrational fundamental of \ce{^80SeC3^80Se} and selected isotopic species (in MHz, unless otherwise noted). Calculated values are set in italics for the sake of clarity.}
	\vspace{.2cm}
	\begin{tabular}{l D{.}{.}{4.11} D{.}{.}{4.11} D{.}{.}{4.11}} \hline \hline
\textbf{Parameter}                & \multicolumn{1}{c}{\textbf{\ce{$^{80}$SeCCCSe$^{80}$}} }           &  \multicolumn{1}{c}{\textbf{\ce{$^{78}$SeCCCSe$^{80}$}}}     & \multicolumn{1}{c}{\textbf{\ce{$^{78}$SeCCCSe$^{78}$}}}\\ 
		\hline
        $\tilde{\nu}_{\rm calc}$ / cm$^{-1,a,b}$      & \textit{2069}.\textit{97}        & \textit{2069}.\textit{92}      & \textit{2069}.\textit{98}   \\
        $\tilde{\nu}_{\rm calc,scaled}$ / cm$^{-1}$   & \textit{n/a}        & \textit{2057}.\textit{16}      & \textit{2057}.\textit{22}      \\
		$\tilde{\nu}_{\rm exp}$ / cm$^{-1}$           & 2057.21101(14) & 2057.23484(10) & 2057.21889(23) \\
		$B_{\rm e}$$^c$                               &  \textit{348}.\textit{574}        &  \textit{352}.\textit{913}       &  \textit{357}.\textit{270}      \\
        $\Delta B_0$$^a$                              &   \textit{0}.\textit{080}        &   \textit{0}.\textit{088}       &   \textit{0}.\textit{090}    \\
        $B_{0,\rm calc}$$^d$                          &  \textit{348}.\textit{487}        &  \textit{352}.\textit{825}       &  \textit{357}.\textit{180}   \\
        $B_{0,\rm calc,scaled}$$^e$                   &   \textit{n/a}            &  \textit{352}.\textit{952}        & \textit{357}.\textit{308}   \\
        $B_0$                                         &  348.605(22)    &  352.918(16)    &  357.201(62) \\
		$\alpha_{3,\rm calc}$$^a$                     &    \textit{1}.\textit{383}        &  \textit{1}.\textit{401}         &    \textit{1}.\textit{418}     \\
        $\alpha_{3}$$^f$                              &    1.3656(8)  &  1.4078(5)    &  1.4678(37) \\	
		$D_{\rm e,calc}\times 10^{6,a}$               &   \textit{2}.\textit{073}          &    \textit{2}.\textit{125}        &  \textit{2}.\textit{177}    \\
		\# lines                                      &\multicolumn{1}{c}{112}&\multicolumn{1}{c}{224}	&\multicolumn{1}{c}{42}\\
		rms / cm$^{-1}$                               &   0.0011         &   0.0012   &    0.0009    \\
        wrms$^g$                                      &   1.11           &   1.16     &    0.92    \\ 
		Abundance                                     & \textit{24}.\textit{8}\:\% & \textit{23}.\textit{6}\:\%  & \textit{5}.\textit{6}\:\%   \\
		Relative Intensity                            & \textit{1}.\textit{0} & \textit{0}.\textit{48} & \textit{0}.\textit{23} \\
		\hline
		\hline
		\\
		\\
\hline \hline
\textbf{Parameter}                & \multicolumn{1}{c}{\textbf{\ce{$^{82}$SeCCCSe$^{80}$}} } &  \multicolumn{1}{c}{\textbf{\ce{$^{77}$SeCCCSe$^{80}$}}} & \multicolumn{1}{c}{\textbf{\ce{$^{76}$SeCCCSe$^{80}$}}}   \\ 
		\hline
        $\tilde{\nu}_{\rm calc}$ / cm$^{-1,a,b}$      &  \textit{2069}.\textit{97}     &  \textit{2069}.\textit{98} & \textit{2069}.\textit{96}       \\
        $\tilde{\nu}_{\rm calc,scaled}$ / cm$^{-1}$   &  \textit{2057}.\textit{20}     & \textit{2057}.\textit{21} & \textit{2057}.\textit{19}       \\
		$\tilde{\nu}_{\rm exp}$ / cm$^{-1}$           & 2057.20463(21) &  2057.20026(23)  & 2057.19919(20) \\
		$B_{\rm e}$$^c$                               &  \textit{344}.\textit{422}        &  \textit{355}.\textit{153}  &  \textit{357}.\textit{454}       \\
        $\Delta B_0$$^a$                              &  \textit{0}.\textit{085}          &  \textit{0}.\textit{089}  &   \textit{0}.\textit{090}       \\
        $B_{0,\rm calc}$$^d$                          & \textit{344}.\textit{337}         &  \textit{355}.\textit{064}  &  \textit{357}.\textit{360}       \\
        $B_{0,\rm calc,scaled}$$^e$                   &  \textit{344}.\textit{461}        & \textit{355}.\textit{191}    &  \textit{357}.\textit{488}       \\
        $B_0$                                         &  344.754(69)& 355.120(67)  &  357.366(101)   \\
		$\alpha_{3,\rm calc}$$^a$                     &   \textit{1}.\textit{366}        & \textit{1}.\textit{410} &  \textit{1}.\textit{419}         \\
        $\alpha_{3}$$^f$                              &    1.3576(40) &  1.4030(42) &  1.4697(53)    \\	
        $D_{e,\rm calc}\times 10^{6,a}$               &  \textit{2}.\textit{024}   & \textit{2}.\textit{152}   & \textit{2}.\textit{180}        \\
		\# lines                                      &\multicolumn{1}{c}{58}&\multicolumn{1}{c}{49}&\multicolumn{1}{c}{58}\\
		rms / cm$^{-1}$                               &   0.0014 &   0.0010   &    0.0009    \\
		wrms$^g$                                      &   1.37   &   1.00     &    0.93    \\
		Abundance                                     & \textit{8}.\textit{7}\:\%   & \textit{7}.\textit{6}\:\%  & \textit{9}.\textit{2}\:\%  \\
		Relative Intensity                            & \textit{0}.\textit{18} & \textit{0}.\textit{15} & \textit{0}.\textit{19} \\
		\hline
		\hline
		
	\end{tabular}
	\vspace{.2cm}
	\begin{minipage}{1.0\textwidth}
		\footnotesize
         \vspace{.2cm}
		$^a$ fc-CCSD(T)/cc-pVTZ.\\
        $^b$ absolute accuracy 0.1 cm$^{-1}$.\\		
        $^c$ CCSD(T)/cc-pwCVTZ. \\
        $^d$ $B_{0,\rm calc}=B_{\rm e}-\Delta B_0$.\\
        $^e$ $X_{\rm calc,scaled} = X_{\rm calc} \times (X_{\rm exp}/X_{\rm calc})_{\rm SeCCCSe}$.\\
        $^f$ $\alpha_{3}\approx B_{0}-B_{3}$.\\
        $^g$ weighted rms, dimensionless. \\
	\end{minipage}
	\label{v3_isotopologs}
\end{table*}

\begin{figure*}[ht!]
\centering
	\includegraphics[width=0.8\textwidth]{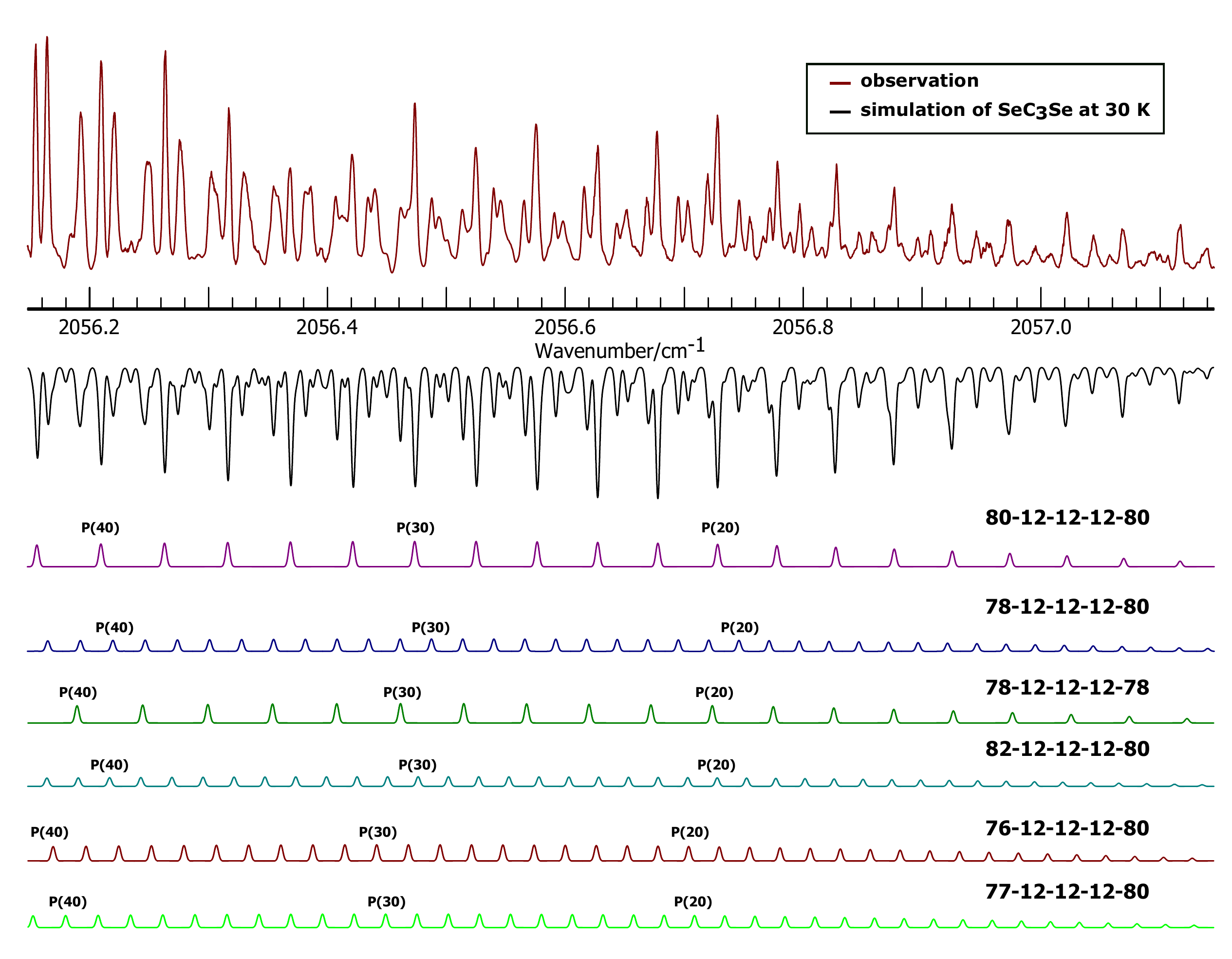}
	\caption[Isotopically shifted modes of \ce{SeC3Se}]{\ce{SeC3Se} $\nu_3$ band close-up at 2056.6\,\wn\ as observed in the present investigation.  
Detail of the  experimental spectrum (maroon trace) obtained at 2056.6\,\wn\ (top) and simulations of the $\nu_3$ fundamentals of six \ce{SeC3Se} species as well as their superposition (black trace). Intensities of individual band simulations are not drawn to scale but to enhance clarity about location in the spectrum. All simulations are based on a rotational temperature of 30K.} 
	\label{se2c3_nu3}
\end{figure*}

Spectroscopic assignment performed in the following commenced with the most abundant and prominent (parent) species, 80--80, and was guided by the calculated rotational constants (Table \ref{v3_isotopologs}) assuming a $4B$ line spacing. 
As the band center region 
at about 2057.2\wn\ is not free of spectroscopic signal and hence the first $R$- and $P$-branch 
lines, $R(0)$ and $P(2)$, were not detectable in a straightforward fashion, the initial assignment was performed by manually adjusting the band origin (while keeping $B_0$ and $B_3$ fixed at their calculated values) until good visual agreement between experiment and simulation was reached.
In a second step, the band origin as well as $B_0$ and $B_3$ were then released in the fitting
procedure and quantum number assignment to the experimental lines was adjusted
such to best reproduce the calculated best-estimate rotational constants (Table \ref{v3_isotopologs}).
Overall, 118 rotational-vibrational transitions -- ranging from P(112) to R(110) -- were assigned to this isotopologue. 
The experimental transition wavenumbers were fit to within experimental accuracy by varying only three parameters using a standard linear rotor
Hamiltonian: the vibrational band center, the ground-state rotational constant $B_0$, and the upper-state rotational constant
$B_3$, or, alternatively, the corresponding rotation-vibration interaction parameter $\alpha_3$. All spectroscopic data were analyzed using Pickett's SPFIT/SPCAT \cite{pickett_JMolSpectrosc_148_371_1991} as well as the Pgopher program \cite{pgopher} (see also the supplementary electronic material).
In the following, despite a considerable number of (partially) overlapping lines and the resulting spectral complexity, 
five additional
isotopic species were identified in the spectrum. Based on isotopic abundance and spin statistical considerations,
the $\nu_3$ band of the 78--80 species is about half as intense as that of the 80--80 species (Table \ref{v3_isotopologs}), followed
by the 78--78 species at about one fourth and the 82--80, 76--80 and 77--80 species
whose bands are expected to be weaker by factors of five to seven. 
Spectroscopic assignment of these species was performed based on
their calculated band centers and lower and upper state rotational constants, $B_0$ and $B_3$,
all of which were scaled further using correction factors derived from a comparison of the calculated and experimental parameters of the parent 80--80
species (see Table \ref{v3_isotopologs}). As in case of \ce{GeC3Ge} \cite{thorwirth_JPCA_120_254_2016}, the vibrational wavenumber of the different isotopic species does hardly depend on the mass of the terminal heavy atoms,
and from the CCSD(T)/cc-pVTZ force-field calculations the bands of all six species were expected to be centered within an interval of less than 0.1\,\wn\ 
(Table \ref{v3_isotopologs}) and this renders theoretical predictions difficult. Despite the resultant spectroscopic interference and diminished line intensities relative to the
parent species, more than 200 transitions were assigned to the 78--80 species, and still a few dozens of lines for
all other species (see Table \ref{v3_isotopologs} and electronic supplementary material).
As can be seen from the final molecular parameter sets summarized in Table \ref{v3_isotopologs}, the empirical scaling
approach is working well and the agreement between the scaled best estimates and experimentally parameters
is good, to (much) better than 0.1\,\wn\ 
for the band centers and some 100\,kHz for the rotational constants.
A 1\,\wn\ snippet of the experimental spectrum at 2056.6\,\wn\ along with a simulation based on the final parameter sets is shown in Figure \ref{se2c3_nu3}. 
Taking account of the isotopic diversity and spectral density
both show very good agreement.

\subsection{Hot bands}

\begin{figure*}[ht!]
\centering
    \includegraphics[width=0.8\textwidth]{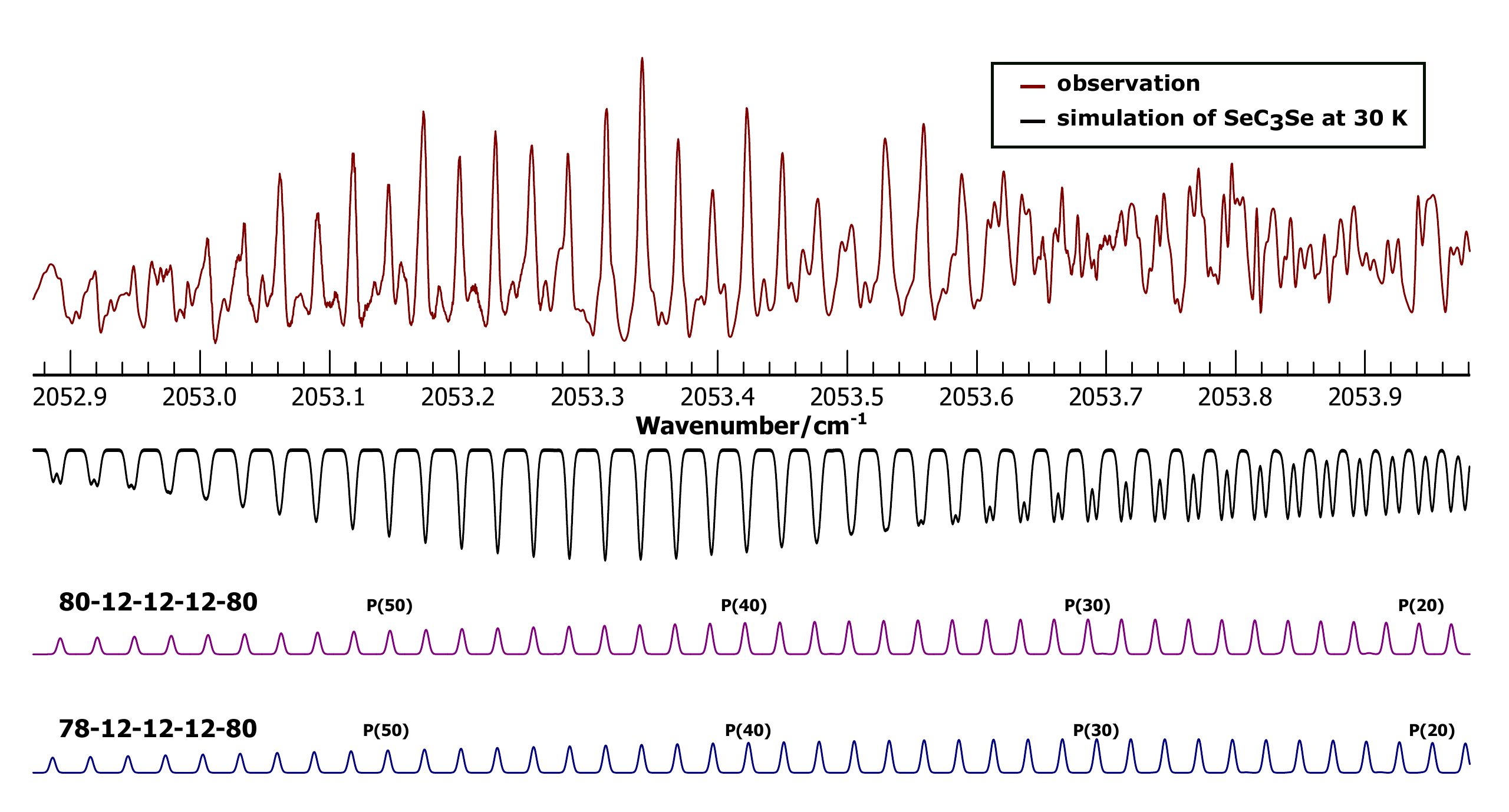}
	\caption[Hot-bands of \ce{SeC3Se}]{The $\nu_3+\nu_5-\nu_5$ hot-bands of \ce{^{80}SeCCC^{80}Se} and \ce{^{78}SeCCC^{80}Se} as observed here vs. a simulation based on the best-fit parameters and a temperature of 30K.}
	\label{hot_bands_se2c3_nu3}
\end{figure*}

The lower-wavenumber band ``tail" visible in Figure \ref{overview_se2c3_nu3} in the 2053 to 2055\,\wn\ region is reminiscent of the band contour also observed in the $\nu_3$ band of SCCCS (see Figure 1 in Ref. \cite{salomon_JMS_356_21_2019}). These weaker spectroscopic features do not stem from any vibrational fundamental of an isotopic species but are hot band transitions associated with the $\nu_3$ modes of the dominant isotopic species, 80--80 and 78--80, and originate from a lower-energy bending vibrational state. The offset 
estimated from simple visual inspection of the spectrum in Figure \ref{overview_se2c3_nu3} is about $-$3\,\wn . 
Based on the calculated 
anharmonicity constants $x_{3j}$ (Table \ref{v3_isotopologs}) the bands are identified as the $\nu_3+\nu_5-\nu_5$ hot bands
and, consequently, the lower state is the first excited $\nu_5$ bending mode (calculated at 352\wn\ for the 80-80 species, Table \ref{qcc_constants}).
As can be deduced from Figure \ref{hot_bands_se2c3_nu3}, spectroscopic assignment in this hot band is not straightforward due to severe line overlaps, a consequence of the overall isotopic richness and most likely also interference with other hot bands (such as $\nu_3+\nu_7-\nu_7$) which could however not be analyzed with confidence here. Guided by the calculated molecular parameters, the comparably clean region between 2053.0 and 2053.6\,\wn\ in which the transitions of the dominating 80--80 and 78--80 species overlap constructively was used in a first assignment procedure and then additional features were added.
Finally, 74 lines of the $\nu_3+\nu_5-\nu_5$ band have been assigned
to the 78--80 species and 127 lines to 80--80.
As the bending modes are doubly degenerate, $\ell$-type doubling will result in a quasi-regular staggered $2B$ spacing in the hot band of the symmetric 80--80 species \cite[cf., e.g, Ref. ][]{neubauer_JChemPhys_127_014313_2007}.
However, the $\ell$-type doubling constant $q_5$ of \ce{SeC3Se} is very small (Table \ref{qcc_constants}) so that line staggering
of adjacent rotational-vibrational transitions is imperceptible in the spectrum. In fact, the effects of staggering are so small, that $q_5$ cannot be determined reliably in the fitting procedure (cf. the situation
in closely related SCCCS \cite{salomon_JMS_356_21_2019}). If released,
$q_5$ amounts to 0.023(47)\,MHz which is in qualitative agreement with the calculation (Table \ref{qcc_constants}).
Consequently, $\ell$-type doubling was also not resolved
for the 78--80 species but both $\ell$-components appear as one single line. 
The finally derived parameters sets are given in Table \ref{tab_hotbands}. 

\subsection{Molecular structure determination}

The rotational constants derived from the observation of six isotopic species
are, in principle, very useful for molecular structure determination. Unfortunately, the present study does not
permit independent ascertainment of both unique structural parameters, $r_{\rm Se-C}$ and $r_{\rm C-C}$. 
Structure determination without constraints
is hampered by the lack of \ce{^13C} data but most severely by the fact
that the moment of inertia
$I=\Sigma_i m_i r_i^2$ of \ce{SeC3Se} is dominated by the heavy masses and the terminal positions of the selenium atoms. As a consequence, $r_{\rm Se-C}$ and $r_{\rm C-C}$ cannot be determined simultaneously in a fully relaxed fitting procedure. 
In analogy to the strategy employed
in case of \ce{GeC3Ge} \cite{thorwirth_JPCA_120_254_2016}, a fit of the carbon-heavy atom
bond length $r_{\rm Se-C}$ was performed (using the equilibrium moments of inertia derived
from $B_e = B_0 + \Delta B_0$, Table \ref{v3_isotopologs})
while keeping $r_{\rm C-C}$ fixed at the theoretical
best-estimate value of 1.2759\,\AA\ calculated at the CCSD(T)cc-pwCVQZ level of theory 
(Figure \ref{se2c3struct}), a level known
to yield highly accurate bond lengths for first-row but also second-row elements \cite[see, e.g., Ref.][]{coriani_JCP_123_184107_2005}. 
From this, an empirical equilibrium bond length $r_{\rm Se-C}$ of 1.6945(3)\,\AA\ is derived (using the STRFIT program \cite{strfit}),
a value being in (virtually quantitative) agreement with the bond length calculated at the CCSD(T)/cc-pwCVQZ level of theory. 
It may be worthwhile to mention that rotational constants derived from alternative spectroscopic assignments ($\pm 2J$, as imposed by the spin statistical constraints of the 80--80 parent species) result in a significant bond length variation $\Delta r_{\rm Se-C}$ of $\pm 0.012$\,\AA , a finding that also speaks very much in favor of the final spectroscopic assignment used in the analysis. 
Assuming a (very conservative) uncertainty in $r_{\rm C-C}$ of 10$^{-3}$\,\AA\ a more conservative
empirical value of $r_{\rm Se-C}=1.695(1)$\,\AA\ is obtained.
Similar to the previous findings on Ge-C chains \cite{thorwirth_JPCA_120_254_2016,lee_PCCP_21_18911_2019}, this result suggests that i) scalar-relativistic effects play a very minor role for the calculation of the Se-C bond length
in \ce{Se2C3} (as deduced also from the CCSD(T)/ANO-RCC calculations highlighted in Figure \ref{se2c3struct}) and ii)
the CCSD(T)/cc-pwCVQZ level may offer a very favorable method-basis set balancing also for molecules harboring third-row main group elements. 

The $r_{\rm Se-C}$ bond length in \ce{SeC3Se} is similar to the one found in linear triatomic OCSe,
for which an empirical equilibrium value of 1.7098\,\AA\ has been determined from combined millimeter-/infrared high resolution spectroscopic studies \cite{leguennec_JMS_157_419_1993}. For OCSe, an CCSD(T)/cc-pwCVQZ structural optimization performed here yields
bond lengths of $r_{\rm O-C}=$1.1529\,\AA\ (vs. an experimental value of 1.1533\,\AA ) and $r_{\rm Se-C}=$1.7108\,\AA . Again, very good agreement between the experimental and calculated carbon-selenium bond lengths is observed, lending independent support as to the 
adequacy of the CCSD(T)/cc-pwCVQZ level for the prediction of high-level structural parameters for carbon-rich selenium clusters in particular and probably even more generally for other selenium bearing species as well. 

\begin{table*}[ht!]
    \scriptsize
	\centering
	\caption{Molecular parameters for the $\nu_3+\nu_5-\nu_5$ vibrational hot-band of \ce{^{80}SeCCC^{80}Se} and \ce{^{78}SeCCC^{80}Se} (in MHz, unless otherwise noted). Calculated values are set in italics for the sake of clarity.}
	\vspace{.2cm}
	\begin{tabular}{ l D{.}{.}{4.11} D{.}{.}{4.11}} \hline \hline
\textbf{Parameter}                & \multicolumn{1}{c}{\textbf{\ce{$^{80}$SeCCCSe$^{80}$}} }           &  \multicolumn{1}{c}{\textbf{\ce{$^{78}$SeCCCSe$^{80}$}}}  \\ 
		\hline

		$\tilde{\nu}_{\rm exp}$/cm$^{-1}$           & 2054.42441(13) &   2054.44142(19)  \\
        $x_{35,\rm calc}^a$/cm$^{-1}$                     & \textit{$-$2}.\textit{77}        &   \textit{$-$2}.\textit{79}             \\			
        $x_{35,\rm exp}$/cm$^{-1}$                          & -2.78660(19)                     &   -2.79342(21)            \\
        $B_{5,\rm calc}$$^b$                          &  \textit{349}.\textit{022}       &   \textit{353}.\textit{335}             \\
        $B_5$                                     &  348.831(40)                     &   353.373(49)       \\
        $\alpha_{3}$$^c$                      &  1.3656(8)                       &   1.4078(5)\\
		$\alpha_{\rm hot}$ $^d$                       &  1.3937(17)                      &   1.4378(30)       \\	
        \# lines                                  &   \multicolumn{1}{c}{127}        &   \multicolumn{1}{c}{74} \\
        rms / cm$^{-1}$                           &   0.0013                         &   0.0012       \\
		wrms$^e$                                      &   1.31                           &   1.23         \\
		\hline
		\hline
	\end{tabular}
	\vspace{.2cm}
	\begin{minipage}{1.0\textwidth}
		\footnotesize
         \vspace{.2cm}
		$^a$ fc-CCSD(T)/cc-pVTZ.\\
        $^b$ $B_{5,\rm calc}=B_{0,\rm exp}-\alpha_{5,\rm calc}$.\\
        $^c$ $\alpha_{3}\approx B_{0}-B_{3}$, see Table 2.\\
        $^d$ $\alpha_{\rm hot}\approx B_{5}-B_{3+5}$.\\
        $^e$ weighted rms, dimensionless.  \\
	\end{minipage}
	\label{tab_hotbands}
\end{table*}

\section{Conclusions}

Laser ablation of carbon-selenium targets has led to the 
first high-resolution spectroscopic study of a polyatomic carbon-selenium cluster, linear \ce{SeC3Se}.
Spectroscopic assignment of the dense $\nu_3$ vibrational mode was made possible and facilitated by
high-level quantum-chemical calculations performed at the CCSD(T) level of theory, despite the overall spectral complexity encountered in the band. Using rotational
constants of six isotopic species permitted the derivation of an accurate carbon-selenium bond length
which is found in very good agreement with the bond length calculated at the CCSD(T)/cc-pwCVQZ level of theory. Scalar-relativistic effects were not found to have a significant impact on the molecular structure.

Based on the present findings, it would be very surprising if there were no other carbon-selenium clusters
present in the laser ablation/free-jet expansion source. Indeed, the $\nu_1$ mode
of the closely related \ce{C3Se} cluster was detected recently in our laboratory \cite{salomon_C3Se_2016} the analysis of which will be described in
detail elsewhere. Longer carbon-selenium chains with more than three carbon atoms building the backbone might be detectable, too,
but enlarging the chain-length will be accompanied by further increase of spectral complexity and line interference making spectroscopic assignment a very challenging task even when predictions from very high level quantum-chemical calculations are at hand.
Ternary chalcogen carbon-rich clusters of the general form OC$_n$Se and SC$_n$Se might also be detectable by similar spectroscopic means in the infrared (and possibly by microwave pure rotational spectroscopy) to open an interesting field for the spectroscopic and structural study of medium-sized and heavy carbon-rich cluster systems.

\section*{Acknowledgments}
We are deeply grateful to Prof. Dr. Stephan Schlemmer for continuous support of this research.
This work has been supported by the Deutsche
Forschungsgemeinschaft (DFG) via SFB 956 (project ID 18401886), DFG SCHL 341/15-1 (“Cologne Center for Terahertz Spectroscopy”), and DFG GA 370/6-2. We thank the Regional Computing Center of the Universit\"at zu K\"oln (RRZK) for providing computing time on the DFG-funded high performing computing system CHEOPS.

\section*{Appendix A. Supplementary material}
Supplementary data associated with this article can be found, in 
the online version, at XXX.



\bibliographystyle{elsarticle-num}         
\bibliography{sthorwirth_bibdesk}           

\end{document}